\documentclass[aps,prl,floatfix,showpacs,twocolumn]{revtex4}
\usepackage{amsmath,amssymb}
\usepackage{epsfig}

\newcommand{\bm}{\boldsymbol}

\begin{document}

%\twocolumn[
\hsize\textwidth\columnwidth\hsize\csname@twocolumnfalse\endcsname

\title{Many-body interaction effects in doped and undoped graphene: \\
Fermi liquid versus non-Fermi liquid}

\author{S. Das Sarma}
\author{E.H. Hwang}
\author{Wang-Kong Tse}
\affiliation{Condensed Matter Theory Center, Department of Physics,
University of Maryland, College Park, Maryland 20742}

\begin{abstract}
We consider 
theoretically the electron-electron interaction induced many-body
effects in undoped (`intrinsic') and doped (`extrinsic') 2D graphene
layers. We find that (1) intrinsic graphene is a \textit{marginal}
Fermi liquid with the imaginary part of the self-energy,
$\mathrm{Im}\Sigma(\omega)$, going as linear in energy $\omega$ for
small $\omega$, implying that the quasiparticle spectral weight
vanishes at the Dirac point as $(\mathrm{ln}\omega)^{-1}$; and, (2)
extrinsic graphene is a well-defined Fermi liquid with
$\mathrm{Im}\Sigma(\omega)\sim \omega^2\mathrm{ln}\omega$ near the
Fermi surface similar to 2D carrier systems with parabolic
energy dispersion. We provide analytical and numerical results for
quasiparticle renormalization in graphene, concluding that all
experimental graphene systems are ordinary 2D Fermi liquids since
any doping automatically induces generic Fermi liquid behavior.
\end{abstract}
%]

\maketitle
\newpage

Recent progress in the experimental realization of a single layer of
graphene \cite{exp1} has spawned tremendous interest and activity in
studying the properties of this unique two-dimensional system.
Graphene has a honeycomb real-space lattice structure, which
comprises two interpenetrating triangular sublattices A and B, and a
concomitant reciprocal space honeycomb structure with the hexagonal
Brillouin zone cornering at the K points. At these so-called ``Dirac
points'' the low-energy excitations satisfy the massless Dirac
equation, and the bandstructure for the ``Dirac fermions'' is
uniquely different from the ``Schr\"odinger fermions'' of the
regular two-dimensional electron gas (2DEG), exhibiting a linear
energy dispersion $\varepsilon_k = vk$ (having an effective ``speed
of light'' $v \simeq 10^6 \mathrm{ms}^{-1}$ \cite{exp1}) with the
conduction and valence bands connected at the Dirac point. 2D
graphene is thus a rather unique chiral Dirac system. This
apparently peculiar bandstructure of graphene is readily accounted
for using a tight-binding model with nearest-neighbour hopping,
where in the vicinity of the crossing of the energy dispersion
relations for the two bands, the energy is linear with respect to
the momentum \cite{Wallace}. In an exciting recent development, this
linear dispersion has been directly observed using angle-resolved
photoemission spectroscopic (ARPES) measurements \cite{Zhou}.
Another recent ARPES experiment \cite{spectral} finds that this
linear spectrum manifests subtle many-body renormalization effects,
preserving, however, the Landau Fermi liquid quasiparticle picture.
On the theoretical side, there has been a number of works in the
literature on the effects of electron-electron interaction in
graphene \cite{Guinea1,Guinea2}; however, the consideration has been
solely based on undoped graphene so far, and explicit analytical
results for either the doped or the undoped case are still lacking. 
The recent ARPES experiments are in doped graphene, and we show in
this paper that doped and undoped graphene have qualitatively
different quasiparticle spectra.

In this paper, we critically address the question of whether
two-dimensional graphene is a Fermi liquid or not using the
diagrammatic perturbation theory. We perform concrete theoretical
calculations for the quasiparticle lifetime, renormalization factor,
and effective velocity for both the doped and the undoped case. We
find that extrinsic graphene manifests a Fermi liquid behavior
similar to the regular 2DEG, indicating that the quasiparticle
description is indeed valid as substantiated by the recent
experiment \cite{spectral}. For intrinsic (i.e. zero doping)
graphene, we find that the quasiparticle lifetime scales linearly
with energy $\omega$ above the Fermi energy (i.e. the Dirac point)
and the renormalization factor at the Fermi energy vanishes,
exhibiting a quintessential marginal Fermi liquid behavior.

Thus, the quasiparticle behavior of 2D graphene depends crucially on
whether the system is doped or not: While undoped intrinsic graphene
is a marginal Fermi liquid, doped graphene with free carriers is
invariably a garden-variety 2D Fermi liquid. Since the presence of
charged impurities in the substrate (i.e. unintentional dopants) would
invariably induce \textit{some} carriers even in nominally undoped
graphene, we conclude that the generic behavior of 2D graphene is
likely to be that of a 2D Fermi liquid.

The low-energy Hamiltonian for graphene is given by $H =
v\bm{\sigma}\cdot\bm{k}$, where $v$ is the Fermi velocity of the Dirac
fermions, $\bm{\sigma}$ is the set of Pauli matrices
representing the two (A and B) sublattice degrees of freedom
(throughout this paper we choose $\hbar = 1$). This Hamiltonian
describes a cone-like linear energy spectrum with conduction band energy
dispersion $vk$ and valence band energy dispersion
$-vk$. The corresponding eigenstates are given respectively by the
plane wave $|\bm{k}\rangle$ multiplied by the spinors
$|\pm\rangle = [1\;\;\,\pm
\mathrm{e}^{i\phi}]/\sqrt{2}$, where $\phi =
\mathrm{tan}^{-1}(k_y/k_x)$ is the polar angle of the momentum $\bm{k}$.
The Green function is given accordingly as $G_{\bm{k}}(ik_n) =
G_{\bm{k}+}(ik_n)|+\rangle\langle+|+G_{\bm{k}-}(ik_n)|-\rangle\langle-|$, where
the label $\pm$, also called the chirality, signifies the conduction band and valence band
respectively, and $G_{\bm{k}\pm}(ik_n) = 1/(ik_n\mp \varepsilon_k)$ is
the Green function in the diagonal basis, where we have denoted
$\varepsilon_k = vk$. The self-energy can also be similarly expressed as
\begin{equation}
\Sigma_{\bm{k}}(ik_n) = \Sigma_{\bm{k}+}(ik_n)|+\rangle\langle+|+\Sigma_{\bm{k}-}(ik_n)|-\rangle\langle-|,
\label{eq1}
\end{equation}
where the leading-order self-energy in the diagonal basis is given by
\begin{eqnarray}
\Sigma_{\bm{k}\pm}(ik_n) &=& -k_{\mathrm{B}}T\sum_{\lambda=\pm}\sum_{\bm{q},iq_n}\left[G_{\bm{k}+\bm{q}\lambda}(ik_n+iq_n)\frac{1\pm\lambda\mathrm{cos}\theta}{2}\right.
  \nonumber \\
&&\left.\;\;\;\;\;\;\;\;\;\;\;\;\;\;\;\;\;\;V_q/\epsilon(q,iq_n)\right],
\label{eq2}
\end{eqnarray}
with the Coulomb potential $V_q = 2\pi e^2/q$, dielectric function
$\epsilon$ and the scattering angle from $\bm{k}$ to $\bm{k}' = \bm{k}+\bm{q}$
denoted as $\theta$. The factor $(1\pm\mathrm{cos}\theta)/2$ comes from
overlap of the eigenstates of the Hamiltonian $H$. Following standard
procedure of analytic continuation, the retarded self-energy from Eq.~(\ref{eq2}) is
obtained as
\begin{eqnarray}
&&\Sigma_{\bm{k}\pm}^{\mathrm{R}}(\varepsilon) =
-\frac{1}{2}\sum_{\lambda =
  \pm}\sum_{\bm{q}}V_{\bm{q}}(1\pm\lambda\mathrm{cos}\theta)\left\{\mathrm{P}\int_{-\infty}^{\infty}\frac{\mathrm{d}\omega}{\pi}n_B(\omega) \right.\nonumber \\
&&\left.\frac{\mathrm{Im}\left[{1}/{\epsilon(q,\omega+i0^+)}\right]}{\varepsilon+\omega-\lambda\varepsilon_{\bm{k}+\bm{q}}+i0^+}+\frac{n_F(\lambda\varepsilon_{\bm{k}+\bm{q}})}{\epsilon(q,\lambda\varepsilon_{\bm{k}+\bm{q}}-\varepsilon-i0^+)}\right\},
\label{eq3}
\end{eqnarray}
where $n_F$ and $n_B$ are the Fermi and Bose distribution functions,
respectively, and $\mathrm{P}$ stands for principal value
integration. In the following, we proceed to calculate the quasiparticle lifetime,
renormalization factor and effective velocity at zero temperature for graphene in two
distinct cases: extrinsic graphene with doping (Fermi energy
$\varepsilon_F > 0$) and intrinsic graphene ($\varepsilon_F =
0$). Note that we calculate our self-energy (Fig.~1) in the
leading-order infinite ring-diagram single-loop expansion
approximation, which is essentially exact in the effective
high-density regime of relevance to graphene. In fact, our self-energy
calculation (Fig.~1) is a much better quantitative approximation to
graphene than the so-called GW approximation in regular metals and
semiconductors which are never in the high-density regime.

\begin{figure}
  \includegraphics[width=8.5cm,angle=0]{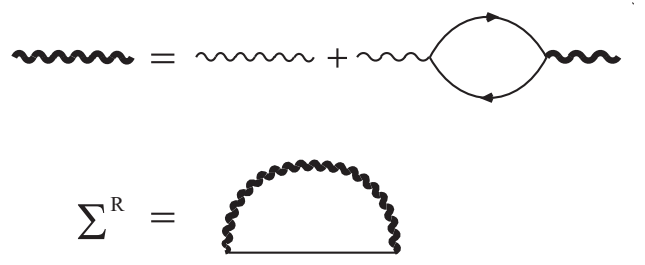}
\caption{The Feynman diagrams for the screened Coulomb interaction
  within the RPA, and the retarded self-energy $\Sigma^{\mathrm{R}}$
  Eq.~(\ref{eq2}). The thin and thick wiggly lines denote respectively the bare and the RPA-screened
  interaction. The thin straight line stands for the bare Green
  function. This is the standard ring diagram approximation for
  self-energy exact in the $r_s \ll 1$ limit.} \label{fig1}
\end{figure}

\textit{Extrinsic graphene.}
For extrinsic graphene, the quasiparticle is located in the vicinity
of the Fermi level and we only need to consider the renormalized Fermi liquid parameters in the conduction
band (or the valence band, depending on whether electrons or holes are
the carriers as determined by doping). The lifetime of the quasiparticle $-1/2\tau_+$ is obtained from the imaginary part
of the self-energy Eq.~(\ref{eq3}) within the on-shell approximation
$\varepsilon = \varepsilon_k$:
\begin{eqnarray}
&&\mathrm{Im}\Sigma_{\bm{k}+}^{\mathrm{R}}(\varepsilon_k) =
\frac{1}{2}\sum_{\lambda =
  \pm}\sum_{\bm{q}}(1+\lambda\mathrm{cos}\theta)V_{\bm{q}}\left[n_B(\lambda\varepsilon_{\bm{k}+\bm{q}}-\varepsilon_k)\right.
\nonumber \\
&&\left.+n_F(\lambda\varepsilon_{\bm{k}+\bm{q}})\right]\mathrm{Im}\left[\frac{1}{\epsilon(q,\lambda\varepsilon_{\bm{k}+\bm{q}}-\varepsilon_k+i0^+)}\right].
\label{eq4}
\end{eqnarray}
At zero temperature, it can be seen that interband scattering from the valence band
$\lambda = -1$ vanishes in Eq.~(\ref{eq4}), and the expression reduces
to (dropping the momentum $\bm{k}$ and energy $\varepsilon$ labels in
the self-energy):
\begin{eqnarray}
\mathrm{Im}\Sigma_{+}^{\mathrm{R}} &=&
-\pi
e^2\int_0^{2\pi}\mathrm{d}\theta(1+\mathrm{cos}\theta)\int_{k_F}^k\mathrm{d}k'\,k'
\nonumber \\
&&\frac{\mathrm{Im}\left[1/\epsilon(|\bm{k}-\bm{k}'|,\varepsilon_{k'}-\varepsilon_k+i0^+)\right]}{\sqrt{k^2+k'^2-2kk'\mathrm{cos}\theta}}.
\label{eq5}
\end{eqnarray}
In order to maintain analytic tractability, we consider the
long-wavelength $x = q/2k_F \ll 1$ limit and perform an analytical evaluation of the quasiparticle
lifetime. For small $x$, the dominant contribution in Eq.~(\ref{eq5}) from
the imaginary part of the dielectric function comes from low
energies $u = \omega/vq \ll 1$, in which case the irreducible
polarizability for graphene (in the doped regime) is given in the leading order by \cite{EH}
$\Pi(q,\omega) \simeq \nu(1+iu)+(q/4v)$, where $\nu = 4\varepsilon_F/2\pi v^2
= 2k_F/\pi v$ is the graphene density of 
states at the Fermi level. 
%While it is recognized that the
%expression for the irreducible polarizability for $x,u \ll 1$ is formally identical to that for the
%regular 2D system with quadratic spectrum,
%we note that for general values of $x$ and $u$ these two expressions are quite different
%\cite{EH}. 
We employ the random phase approximation (RPA) (Fig.~\ref{fig1}) for
the screened Coulomb potential and calculate the dielectric function
as $\epsilon(q,\omega)/\epsilon_m = 1+V_{q}\Pi(q,\omega)$ (here $\epsilon_m$ is the background dielectric
constant).
Substituting this expression for the dielectric function in Eq.~(\ref{eq5}), we obtain
\begin{eqnarray}
&&\mathrm{Im}\Sigma_{+}^{\mathrm{R}} = \label{eq6} \\
&&\frac{\pi e^2}{q_{\mathrm{TF}}}\int_{k_F}^k\mathrm{d}k'\,k'(k-k')\int_0^{2\pi}\mathrm{d}\theta\frac{(1+\mathrm{cos}\theta)}{{\sqrt{k^2+k'^2-2kk'\mathrm{cos}\theta}}},
\nonumber
\end{eqnarray}
where we have defined the Thomas-Fermi wavenumber as $q_{\mathrm{TF}}
= 4e^2k_F/v\epsilon_m$. After carrying out the angular integral, expanding in
the small parameter $\delta = k-k' \ll k_F$, and integrating over $k'$,
we obtain
\begin{eqnarray}
\mathrm{Im}\Sigma_{+}^{\mathrm{R}} = \frac{\xi_k^2}{8\pi\varepsilon_F}\left[\mathrm{ln}\left(\frac{\xi_k}{8\varepsilon_F}\right)+\frac{1}{2}\right],
\label{eq7}
\end{eqnarray}
where $\xi_k = \varepsilon_k-\varepsilon_F$ is the single-particle energy reckoned from
the Fermi level. Comparing with the corresponding expression for the
regular 2DEG \cite{LZheng} $\mathrm{Im}\Sigma^{\mathrm{R}} =
({\xi_k^2}/{2\pi\varepsilon_F})[\mathrm{ln}({\xi_k}/{16\varepsilon_F})-{1}/{2}]$,
it is instructive to note that (1) the factor of two difference
inside the logarithm comes from the linearity of the graphene
spectrum; (2) the sign difference in the subleading term comes from
the vanishing of the interband contribution from the valence band in
Eq.~(\ref{eq4}) at zero temperature.

Next we consider the renormalization factor for the Dirac
quasiparticle $Z = 1/(1-A)$, where $A$ is the derivative of the real
part of the self-energy Eq.~(\ref{eq2}) with respect to
energy:
\begin{eqnarray}
A &=&
-\frac{1}{2}\frac{\partial}{\partial\varepsilon}\mathrm{Re}\sum_{\bm{q},\lambda}\int_{-\infty}^{\infty}\frac{\mathrm{d}\omega}{2\pi}
G_{\bm{k}+\bm{q}\lambda}(ik_n+i\omega) \nonumber \\
&&(1+\lambda\mathrm{cos}\theta)V_{\bm{q}}/\epsilon(q,i\omega)\big\vert_{k,\varepsilon = k_F,\varepsilon_F}.
\label{eq8}
\end{eqnarray}
As usual, we perform the standard trick for evaluating the real part
of the retarded self-energy by decomposing Eq.~(\ref{eq8}) into line
and pole contributions \cite{Mahan} and performing analytic
continuation $ik_n \to \varepsilon+i0^+$, and then integrating by
parts the $\omega$-integral \cite{Saraga}, obtaining
\begin{eqnarray}
A &=&
\frac{1}{\pi}\mathrm{Im}\sum_{\bm{q},\lambda}\int_0^{\infty}\mathrm{d}\omega\,
G_{\bm{k}+\bm{q}\lambda}(i\omega)(1+\lambda\mathrm{cos}\theta)V_{\bm{q}}
\nonumber \\
&&\frac{\partial}{\partial
  \omega}\frac{1}{\epsilon(q,i\omega)}\bigg\vert_{k = k_F}.
\label{eq9}
\end{eqnarray}
%
%In evaluating the $q$-integral Eq.~(\ref{eq9}), 
%we keepwhile for the interband term  
%taking the Green function as 
%$G_{\bm{k}+\bm{q}\lambda}(i\omega) \simeq
%(1/vq)/(iu-\lambda\mathrm{cos}\phi)$ and 
The irreducible polarizability \cite{EH} is 
$\Pi(q,\omega) \simeq \nu(1-u/\sqrt{u^2-1})+i(q/4v)/\sqrt{u^2-1}$. 
%; in addition, 
%we neglect the interband contribution as it is of an order $\mathcal{O}(x^2)$ smaller than the intraband
%contribution. 
After some straightforward algebra, we obtain the
renormalization factor for the Dirac quasiparticle in the $r_s \ll 1$ limit as
\begin{equation}
Z = 1/\left[1+\frac{r_s}{\pi}\left(1+\frac{\pi}{2}\right)\right],
\label{eq10}
\end{equation}
where we have defined the interaction parameter $r_s$ by the ratio
of the interparticle potential energy to the single-particle kinetic
energy. It is interesting to note that whereas for regular 2DEG, $r_s =
2me^2/k_F\epsilon_m$ is inversely proportional to the square root of
the 2D density, for graphene $r_s = e^2/v\epsilon_m$ is simply a
constant $\sim 0.73$ (with $v \simeq 10^6\mathrm{ms}^{-1}$ and
$\epsilon_m \simeq 3$), indicating extrinsic graphene is essentially a
weakly interacting ($r_s < 1$) system. (We note in this context that
ordinary metals have $r_s \simeq 3-5 > 1$, and the usual
semiconductor-based 2DEG may have $r_s \sim 5-20 \gg 1$\,!) In the leading order of $r_s$, Eq.~(\ref{eq10}) is similar 
to the corresponding expression for regular 2DEG \cite{Burkard} $Z = 
1-(r_s/2\pi)(1+\pi/2)$. Beyond the leading order of $r_s$, the $q$-integral in $A$ has a logarithmic divergence 
due to the interband contribution to the polarizability. Introducing a momentum cutoff $k_c$ of the order of inverse lattice spacing, we find the corresponding logarithmic correction $\sim r_s^2\mathrm{ln}(k_c/k_F)$ appears in the second order of $r_s$ in the denominator of Eq.~(\ref{eq10}).  
%The difference in the prefactor of $r_s$ in
%graphene is due to a combination of numerical factors coming from
%the extra spin and valley degeneracies and the linearity of the
%energy spectrum (i.e. $\varepsilon_k = vk$ instead of $k^2/2m$). 

For quasiparticles with regular quadratic spectrum,
it is well-known \cite{Mahan} that electron-electron interaction gives rise
to effective mass renormalization of the Fermi liquid. In contrast, as the Dirac
quasiparticle in graphene is massless, we have, instead, a
renormalization of the quasiparticle velocity. We proceed to calculate
the effective velocity renormalization below, which is defined through
$v/v^* = (1-A)/(1+B)$, where $B$ is the derivative of the real part of
the retarded self-energy with respect to momentum $k$ and is obtained
from Eq.~(\ref{eq8}) by replacing ${\partial}/{\partial
  \varepsilon} \to (1/v){\partial}/{\partial k}$.
Decomposing the expression for $B$ into line and pole contributions and taking the analytic
continuation $ik_n \to \varepsilon+i0^+$, we find that the pole contribution
vanishes and the expression for $B$ becomes
\begin{eqnarray}
B &=&
-\frac{1}{2v}\mathrm{Re}\sum_{\bm{q},\lambda}\int_0^{\infty}\frac{\mathrm{d}\omega}{\pi}
\frac{\partial}{\partial k}\left[G_{\bm{k}+\bm{q}\lambda}(ik_n+i\omega)\right. \nonumber \\
&&\left.(1+\lambda\mathrm{cos}\theta)V_{\bm{q}}/\epsilon(q,i\omega)\right]\big\vert_{k,\varepsilon = k_F,\varepsilon_F},
\label{eq11}
\end{eqnarray}
The evaluation of Eq.~(\ref{eq11}) is similar to that for the
renormalization factor Eq.~(\ref{eq9}), albeit more tedious.
Carrying out the $\omega$-integral by integration by parts \cite{Saraga}, and then
performing the $q$-integral, we obtain 
%
%\begin{equation}
%B =
%-\frac{r_s}{\pi}\left[\frac{2}{3}+\mathrm{ln}(r_s)-\frac{\pi}{2}\right],
%\label{eq12}
%\end{equation}
%
the renormalized velocity at the Fermi level, within logarithmic accuracy, as 
\begin{equation}
\frac{v^*}{v} = 1-\frac{r_s}{\pi}\left[\frac{5}{3}+\mathrm{ln}(r_s)\right]+\frac{r_s}{4}\mathrm{ln}(\frac{k_c}{k_F}).
\label{eq13}
\end{equation}
%
%where $k_c$ is a cutoff momentum of the order of inverse lattice spacing introduced 
%to make the interband contribution finite.  
We also note that the same result here can be obtained more simply by using the static
dielectric function and taking the derivative of the quasiparticle
energy, consistent with the fact that the effective velocity at zero
temperature in the lowest leading order in $r_s$ is only due to contribution from the
static dielectric response. The first two terms in Eq.~(\ref{eq13}) 
derive from the intraband contribution and are similar to 
the expression for the regular 2DEG \cite{Janak} $m^*/m = 1+(r_s/2\pi)[2+\mathrm{ln}(r_s/4)]$ 
%(besides the difference in the prefactors arising from degeneracy and
%linear spectrum), 
whereas the last term arises solely from the interband contribution. 
In addition to $v^*$ at the Fermi level, we have 
evaluated $v^*$ at the Dirac point $k = 0$:  
$v^*/v = 1-r_s\{1+(1/4)\mathrm{ln}[(1+4r_s)/4r_s]-(1/4)\mathrm{ln}(k_c/k_F)\}$, 
highlighting the fact that velocity renormalization is not uniform for 
the entire spectrum but is in general a function of $k$. The 
renormalized spectrum therefore exhibits a small degree of 
nonlinearity imposed on the bare linear spectrum.    
%The renormalized velocity is in general a function of $k$, and in
%addition to $v^*$ at the Fermi level, 
%we have evaluated $v^*$ exactly at the Dirac point $k = 0$:   
%$v^*/v = 1-(r_s/2)\{\mathrm{ln}[(1+4r_s)/4r_s]-1/(1+4r_s)\}$, 
%highlighting the fact that the velocity renormalization at different
%points of the linear spectrum is generally different. 
%

\textit{Intrinsic graphene.}
The irreducible polarizability for intrinsic graphene is given by \cite{EH}
\begin{equation}
\Pi(q,\omega) = \frac{q^2}{4}\left[\frac{\theta(vq-\omega)}{\sqrt{v^2 q^2-\omega^2}}+i\frac{\theta(\omega-vq)}{\sqrt{\omega^2-v^2 q^2}}\right].
\label{eq14}
\end{equation}
Using Eq.~(\ref{eq14}), the on-shell quasiparticle lifetime follows from
Eq.~(\ref{eq4}) as $\mathrm{Im}\Sigma_{\bm{k}+}^{\mathrm{R}}(\varepsilon_k) \sim
\sum_{\bm{q}}\theta(|\varepsilon_{\bm{k}+\bm{q}}-\varepsilon_k|-\varepsilon_q)\theta(\varepsilon_k-\varepsilon_{\bm{k}+\bm{q}})$,
which vanishes identically because of phase space restrictions imposed
by the $\theta$-functions. On the other hand, the imaginary part of
the self-energy at $k=0$ follows a linear relationship with $\omega$,
and is given by
\begin{equation}
\mathrm{Im}\Sigma^R_{\pm} (k=0,\omega) = \omega f(r_s),
\label{eq15}
\end{equation}
where
\begin{eqnarray}
f(r_s) & = & \frac{2}{\pi^2 r_s}
\left [\pi(1-r_s) \right . \nonumber \\
& + &
\left .  \frac{8-(\pi r_s)^2}{4\sqrt{(\pi r_s)^2-4}}\ln\frac{\pi r_s -
    \sqrt{(\pi r_s)^2-4}}{\pi r_s + \sqrt{(\pi r_s)^2-4}}
\right ].
\label{eq16}
\end{eqnarray}
We emphasize that the linear relation is exact for all $\omega$,
indicating that intrinsic graphene is a marginal Fermi liquid. For
arbitrary values of $k$ we calculate
$\mathrm{Im}\Sigma_{\pm}^{\mathrm{R}}(k,\omega)$ numerically, and
Fig.~2 shows our calculated self-energy as a function of energy
$\omega$ for various values of momentum $k$. The numerical result
for $k = 0$ in Fig. 2 agrees precisely with the analytic result of
Eq.~(\ref{eq16}), providing a consistency check. As mentioned above
we have Im$\Sigma_{\pm}^{\mathrm{R}} =0$ for $\omega \le
\varepsilon_k$ because there is no phase space available for virtual
interband electron-hole excitations. However, for $\omega >
\varepsilon_k$ the virtual interband electron-hole excitations give
rise to a finite Im$\Sigma_{\pm}^{\mathrm{R}}$ (i.e. finite
quasiparticle lifetime). Initially Im$\Sigma_+^{\mathrm{R}}$
(Im$\Sigma_-^{\mathrm{R}}$) rises sharply (slowly), and for large
values of momentum they increase linearly with the same slope as
that for $\mathrm{Im}\Sigma_{\pm}^{\mathrm{R}}(k = 0,\omega)$ (c.f.
Eq.~(\ref{eq16})). Note that there is no plasmon contribution to the
imaginary part of the self-energy for intrinsic graphene. The
contribution of the interband electron-hole excitations gives rise
to the linear behavior of $\mathrm{Im}\Sigma_{\pm}^{\mathrm{R}}$.
However, for doped graphene the contributions of the interband
electron-hole excitations is completely suppressed due to phase
space restrictions at zero temperature. The contributions of the
intraband virtual single-particle excitations and/or the virtual
excitations of plasmons give rise to higher powers of $\omega$ (i.e.
$\omega^2$) in the imaginary part of the self-energy in the doped
case, which restores the usual Fermi liquid behaviour. Thus, the
\textit{qualitative} difference between intrinsic
($\mathrm{Im}\Sigma \sim \omega)$) and extrinsic ($\mathrm{Im}\Sigma
\sim \omega^2$) graphene can be \textit{completely} understood by
noting that the intrinsic system is an \textit{insulator} (albeit a
zero-gap semiconductor with no intraband single-particle excitation)
and the extrinsic case has a Fermi surface with intraband
single-particle excitations. Thus, any doping of graphene
(intentional or unintentional) will immediately suppress its
marginal Fermi liquid intrinsic character, converting it to a
regular 2D Fermi liquid.
\begin{figure}
%\epsfysize=1.8in
%\centerline{\epsffile{Im_self.eps}}
%\vspace{0.5cm}
\includegraphics[width=8.5cm,angle=0]{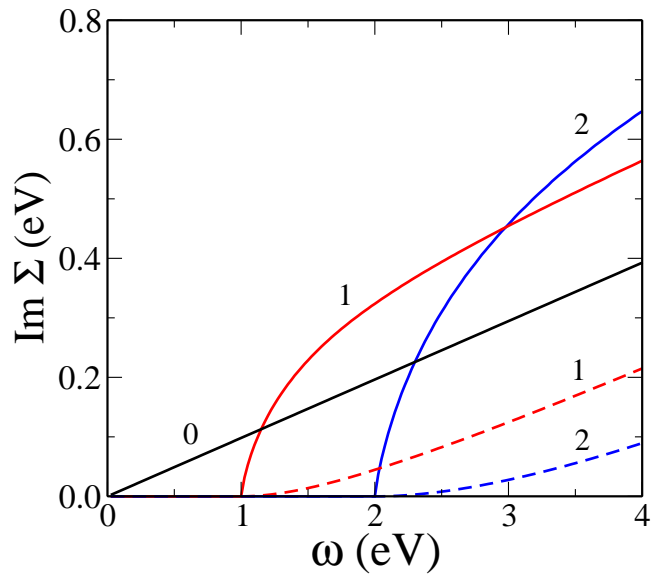}
\caption{(Color online) Imaginary part of the self-energy for the intrinsic graphene,
$\mathrm{Im}\Sigma^{\mathrm{R}}_{+}(k,\omega)$ (solid lines),
$\mathrm{Im}\Sigma^{\mathrm{R}}_-(k,\omega)$
(dashed lines), for different values of momentum, $k=0,\,1,\,2\;\mathrm{eV}/\hbar v$
(black line `$0$', red line `$1$', blue line `$2$', respectively). Note that for $k=0$, $\mathrm{Im}\Sigma^{\mathrm{R}}_+
= \mathrm{Im}\Sigma^{\mathrm{R}}_-$.
}
\end{figure}

The renormalization factor can be obtained from Eq.~(\ref{eq9}),
whereupon evaluating we find that $A \sim \int_0 \mathrm{d}q/q$
diverges logarithmically, which is due to the unscreened nature of
the Coulomb potential (the same divergence occurs for the exchange
energy of a regular 2DEG). Therefore the renormalization factor $Z =
0$, showing that as doping goes to zero, the magnitude of the step
at the Fermi energy $\varepsilon = 0$ also shrinks to zero,
approaching the Dirac point where the notion of a ``Fermi surface''
no longer applies, the quintessential behaviour of a marginal Fermi
liquid. Using the Kramers-Kronig relations and from the fact that
$\mathrm{Im}\Sigma_{+}^{\mathrm{R}}(k = 0,\omega) \sim \omega$, we
have $\mathrm{Re}\Sigma_{+}^{\mathrm{R}}(k = 0,\omega) \sim
\omega\,\mathrm{ln}\omega$, and the renormalization factor $Z \sim
1/\mathrm{ln}\omega$, which approaches zero logarithmically as
$\omega \to 0$ at the Fermi energy. Moreover, the spectral function
$\rho(k = 0,\omega) = \mathrm{Im}\Sigma_+^R(k = 0,\omega)/\{
[\mathrm{Im}\Sigma_+^R(k =
0,\omega)]^2+[\omega-\mathrm{Re}\Sigma_+^R(k = 0,\omega)]^2\}$
diverges as $\rho(k = 0,\omega) \sim
1/\omega\,(\mathrm{ln}\omega)^2$. In addition, we also find that the effective velocity 
$v^*/v = 1+(r_s^*/4)\mathrm{ln}(k_c/k_F)$ (here $r_s^* = r_s/(1+\pi r_s/2)$ is the renormalized interaction parameter) diverges as doping $k_F \to 0$.
These results for the intrinsic
graphene are consistent with Ref.~\cite{Guinea2}, where the
renormalization group approach is used to arrive at a similar
conclusion. We note in passing that the case for a purely undoped 3D system with a gapless linear 
energy dispersion was considered in Ref.~\cite{Absov}, and was found to 
exhibit marginal Fermi liquid behavior with a logarithmic 
energy dependence in $\mathrm{Re}\Sigma$ comparable to our results for
intrinsic graphene.  

In conclusion, we have presented a calculation, formally exact in
the $r_s \ll 1$ limit, for the renormalized Fermi liquid parameters
for both extrinsic and intrinsic graphene. We find that for
extrinsic graphene the analytical results for the quasiparticle
lifetime, renormalization factor and effective velocity show no
deviation from the usual Fermi liquid behaviour, and the Fermi
liquid description is robust. On the other hand, with precise zero
doping, intrinsic graphene exhibits a quasiparticle lifetime linear
in the excitation energy and a zero renormalization factor,
indicating that the Fermi liquid description is marginal at the
Dirac point. With a finite Fermi energy in the extrinsic graphene,
the interband single-particle excitations which give rise to the
linear $\omega$-dependence of the quasiparticle lifetime (and hence
the marginal Fermi liquid behavior) in the intrinsic graphene are
suppressed, bringing the system back to a usual Fermi liquid. Since
some finite doping is invariable in real systems, we predict real 2D
graphene to be generically a Fermi liquid. 

This work is supported by US-ONR.

\end{document}